\documentclass[unnumsec,webpdf,contemporary,large]{oup-authoring-template}
\graphicspath{{Fig/}}
\usepackage{graphicx} 

\begin{document}

\journaltitle{arxiv}
\DOI{N/A}
\copyrightyear{2024}
\pubyear{2024}
\access{arxiv openaccess}
\appnotes{application}
\application

\firstpage{1}

\title[]{TIAViz: A Browser-based Visualization Tool for Computational Pathology Models}
\author[1,$\ast$]{Mark Eastwood}
\author[3]{John Pocock}
\author[1]{Mostafa Jahanifar}
\author[1]{Adam Shephard}
\author[1]{Skiros Habib}
\author[1]{Ethar Alzaid}
\author[1]{Abdullah Alsalemi}
\author[2]{Jan Lukas Robertus}
\author[1,3]{Nasir Rajpoot}
\author[1]{Shan Raza}
\author[1]{Fayyaz Minhas}

\authormark{Eastwood et al.}

\address[1]{\orgdiv{Tissue Image Analytics Center}, \orgname{University of Warwick}, \orgaddress{\country{UK}}}

\address[2]{\orgdiv{National Heart and Lung Institute}, \orgname{Imperial College London}, \orgaddress{\country{UK}}}

\address[3]{\orgname{Histofy}, \orgaddress{\country{UK}}}

\corresp[$\ast$]{Corresponding author. \href{email:email-id.com}{Mark.Eastwood@warwick.ac.uk}}

\abstract{
\absection{Summary} 
Digital pathology has gained significant traction in modern healthcare systems. This shift from optical microscopes to digital imagery brings with it the potential for improved diagnosis, efficiency, and the integration of AI tools into the pathologists workflow. A critical aspect of this is visualization. Throughout the development of a machine learning (ML) model in digital pathology, it is crucial to have flexible, openly available tools to visualize models, from their outputs and predictions to the underlying annotations and images used to train or test a model. We introduce TIAViz, a Python-based visualization tool built into TIAToolbox which allows flexible, interactive, fully zoomable overlay of a wide variety of information onto whole slide images, including graphs, heatmaps, segmentations, annotations and other WSIs. The UI is browser-based, allowing use either locally, on a remote machine, or on a server to provide publicly available demos.
\absection{Availability and Implementation} 
This tool is open source and is made available at: \url{https://github.com/TissueImageAnalytics/tiatoolbox} and via pip installation (pip install tiatoolbox) and conda as part of TIAToolbox.
\absection{Contact} 
For any queries please contact Mark.Eastwood@warwick.ac.uk, or raise an issue in the associated git repository above.
\absection{Supplementary information} 
Further documentation can be found at: \url{https://tia-toolbox.readthedocs.io/en/latest/visualization.html}. Some demos that use the described tool can be found here: \url{tiademos.dcs.warwick.ac.uk}.
}

\maketitle

\section{Introduction}

In bioinformatics and digital pathology, the need for advanced visualization tools is driven by the growing complexity and prevalence of AI models capable of providing additional context and predictions to pathologists when viewing slides. 

TIAViz is a browser-based visualization tool designed for the analysis of digital pathology model outputs. Developed as part of the TIAToolbox \cite{pocock2022tiatoolbox} open-source digital pathology toolbox, and utilizing the Bokeh \cite{Bokeh} visualization library, it provides a platform for viewing high-resolution pathology slides and overlaying model outputs or annotations in a fully zoomable and interactive viewer. This tool helps address the increasing demand in digital pathology for robust and flexible visualization options capable of handling complex medical imaging data and model outputs in a variety of different forms, with a particular focus on use within a research context.
As all code is written in Python it can be integrated quickly into Python-based machine learning workflows.

There are some excellent open source tools for digital pathology already available. The foremost of these is probably QuPath \cite{qupath2017}, a fully featured, java-based platform for analysis of histological images. For many use-cases, we would highly recommend QuPath, such as for creating and working with annotations, interactive pixel-level analysis and modelling. However, there are a few areas in which TIAViz can offer functionality that is difficult or not provided in QuPath. Firstly, 
as a browser-based visualization tool, TIAViz can be launched on a remote machine and used to view slides/overlays from that remote machine in a browser window on the local machine. This can be very useful, especially in the research context as ML pipelines and model
outputs are often run and stored on remote compute resources.

Additionally, it's browser-based nature allows it to be used to provide online servers showcasing the capabilities of models, which can be extremely useful for collaboration and to allow members of the DP (Digital Pathology) community to better understand and explore models created by others.

TIAViz is written in Python, using TIAToolbox \cite{pocock2022tiatoolbox} and Bokeh \cite{Bokeh}. As such, it can be easier to integrate into Python-based ML workflows, in particular those that already use elements of TIAToolbox, and has a much simpler code-base allowing it to potentially be adapted or customized quite easily to a specific use case when the desired visualization task cannot be achieved with the tool's inbuilt capabilities.

Another tool is cytomine \url{https://cytomine.com/}, which is browser-based but is limited in what can be overlaid (for example graphs are not supported). Similarly, HistomicsTK (\url{https://github.com/DigitalSlideArchive/HistomicsTK}) is an option with a powerful set of tools for annotation creation, management, and image analysis, but again graphs are not supported and it is less well suited to quickly visualizing ML model output.

\section{Technical Details}

TIAToolbox is a suite of unit-tested image analysis and machine learning tools developed for the CPath community, making it possible for a variety of users to construct and reproduce CPath analytical pipelines with cutting-edge methods. TIAViz builds on the capabilities of TIAToolbox, utilizing two core components of the package. Firstly, TIAToolbox provides the capability to read a wide variety of image formats, and a flask-based tileserver serves image tiles on the fly to allow for responsive navigation around large WSIs.

Secondly, TIAToolbox offers an SQLite-based database for storing annotations, as geometries paired with a properties dictionary. Methods are provided to allow efficient and powerful querying of the store, to select annotations by some combination of spatial location or conditions on their property values.

We add significantly to the tileserver and annotation store, and introduce a variety of classes and functions to facilitate rendering of annotation overlay tiles directly from an annotation store, and for conversion of some common annotation types to an AnnotationStore database. 

The interface itself is developed using the Bokeh visualization library, a powerful Python toolkit for creating interactive plots and dashboards. This choice of technology facilitates dynamic data visualization and interactivity, essential for handling the complex data typical in digital pathology. 

\section{Features and Capabilities}

\begin{figure*}[hp]
\centering
\centerline{\includegraphics[width=16.8cm]{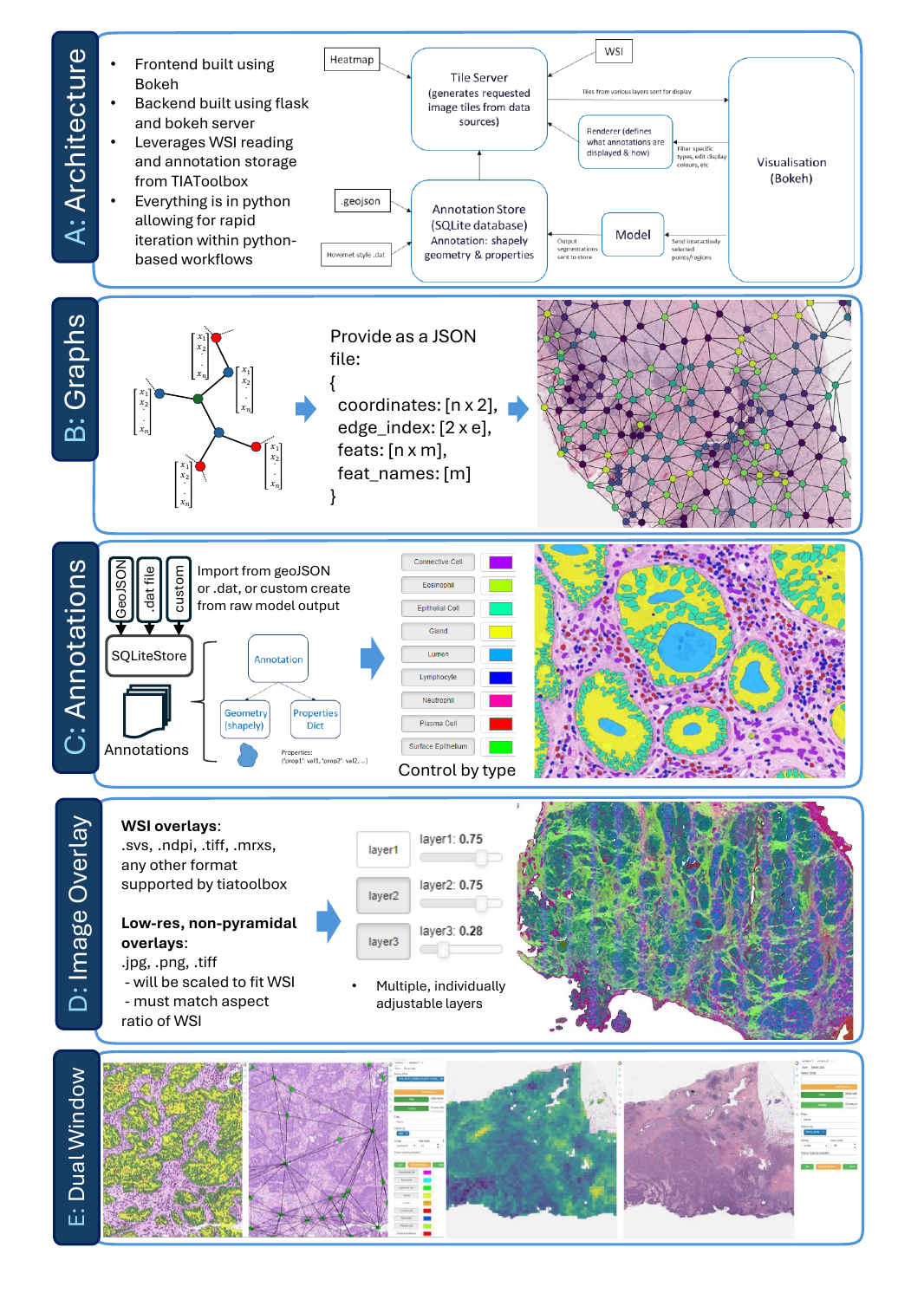}}
\caption{Overview of TIAViz architecture and capabilities. Panel A: High level architecture and data flow. B: Graph overlay features. C: Overlay of annotations and segmentation. D: Image-based overlays. E: Dual linked window capabilities.
\label{overview_fig}}
\end{figure*}

TIAViz is capable of overlaying a wide variety of model outputs and other information onto a WSI. A straightforward file structure is assumed, requiring all slides to be in a designated `slides' folder, and all overlays - including graphs, segmentations, heatmaps, etc. - in an `overlays' folder. The tool automatically pairs overlays with slides based on if an overlay filename stem contains the slide stem, streamlining the process of data visualization.

\subsection{Annotation and Segmentation Overlays}
The tool is capable of overlaying geometries provided in various formats. TIAViz uses the TIAToolbox SQLiteStore to store annotations and segmentations as Annotation objects, which consist of a geometry and any associated properties. GeoJSON files, and .dat file output from TIAtoolbox segmentation models, can also be loaded in the interface.

We provide a powerful selection of UI elements to interactively control how annotations are displayed, allowing individual types to be toggled on/off independently, filtering of annotations according to mathematical expressions, and offering a wide variety of options to allow coloring annotations according to any values in their properties dictionary. Detailed annotation properties can be viewed by double-clicking on an annotation, as illustrated in supplementary Fig 2.

\subsection{Heatmaps and Whole Slide Overlays}
TIAViz supports the overlay of multiple low-resolution heatmaps, and/or additional WSI overlays each as separate layers. Low-res images must be provided with the same aspect ratio as the parent image, and will be upscaled accordingly. An example of a UMAP \cite{McInnes2018} reduction of a swin-transformer \cite{liu2021swin} featuremap into 3-D (RGB) space is shown in \ref{overview_fig} Panel C.

\subsection{Graph Overlays}
Graphs are becoming an increasingly popular way of representing large sections of tissue in histology images, paired with graph neural network (GNN) models for prediction \cite{ahmedt2022survey}. TIAViz allows graphs to be overlaid onto WSIs in tandem with other overlay types, and can color nodes according to provided features or model output. Users can control the display of nodes and edges individually. The tool accepts graph data in a JSON dictionary format. Examples of this can be seen in \ref{overview_fig}. Futher information about node features can be viewed when hovering over a node, as illustrated in supplementary Fig 3.

\subsection{Interactive AI Environment}
Models can be run interactively in the UI. HoVer-Net \cite{graham2019hover} is the first model available this way and can be used to segment small regions of a slide interactively.
GPT-vision has been shown to have some ability to evaluate medical images \cite{yang2023gptv}. TIAViz provides the ability to highlight a region of a slide and send it to GPT-vision within the interface. Drawn annotations (for example circling a feature of interest) can be included in the sent image, and referenced in the prompt. Examples can be seen in supplementary Fig 1. We believe this sort of functionality could be very powerful in the future with multi-modal models fine-tuned for medical images, as it would allow pathologists to interact intuitively in natural language with an AI assistant. Code for this addition is available in a fork of TIAViz at \url{https://github.com/measty/tiatoolbox}.

\subsection{Other Features}
A powerful feature of this tool is the ability to display multiple overlays simultaneously. Any combination of one graph overlay, one annotation-based overlay, and multiple image-based overlays (either WSIs or low resolution non-pyramidal images) can be shown.

It is also possible to display two slides (or two views of the same slide) side by side, and add overlays to each independently, as shown in \ref{overview_fig}, Panel E. Slide navigation will be linked between the two windows. This is extremely useful for viewing registered slides, comparing different model output for the same slide, or simply showing the base image next to an overlaid image.

A JSON configuration file can be provided which can define a variety of defaults for a given project, such as default colors for specific types. More information can be found in the documentation \url{https://tia-toolbox.readthedocs.io/en/latest/visualization.html}.

\section{Example Applications}

The TIAToolbox visualization tool has been utilized in a range of research works in digital pathology, some of which we will cover here to illustrate its flexibility.

\textbf{IGUANA}:  This is an interpretable graph neural network for screening normal endoscopic large bowel biopsies \cite{graham2023screening}. The pipeline for this model involves the segmentation of glands and several different cell types, which are used to build a gland graph with interpretable node features calculated from gland and nearby cell segmentations. These features were developed in collaboration with pathologists to capture aspects of glandular morphology and the surrounding lamina propria that they consider when making their diagnosis. TIAViz is highly suitable for visualizing such information, providing the ability to simultaneously overlay gland and cell segmentations together with the gland graph, and visually highlight the glands determined by the model to contribute most highly to an abnormal diagnosis. An online demo illustrating the IGUANA model can be found at \url{https://iguana.dcs.warwick.ac.uk/}.  

\textbf{Social Network Analysis (SNA)}: In \cite{neda2023sna}, SNA is used to characterize patches of WSI based on the relative locations of cells in the tissue. These features are used to create an enhanced IDaRS-based model \cite{idars2021} to predict molecular pathways and mutations in colorectal cancer. The task is treated as a Multiple instance learning problem. TIAViz provides a convenient way to visualize the most positive instances (patches) on a WSI, as can be seen in \url{https://tiademos.dcs.warwick.ac.uk/Bokeh_app?demo=sna}.

\textbf{Gene group analysis}: Work in \cite{dawood2023data} has identified groups of genes with co-dependent expression and predicted their status from WSI images using a graph neural network. These gene groups allow to capture the gene expression state of a patient with a small number of binary variables that are biologically meaningful. Further, the  groups can be used to associate histological phenotypes (cellular composition, mitotic counts, grading, etc.) with underlying gene expression patterns. Our visualization tool provides a convenient interface to explore the predicted expression of the gene groups and visualize their variation and correlation with visual features over a WSI, as illustrated at \url{https://tiademos.dcs.warwick.ac.uk/Bokeh_app?demo=HiGGsXplore}.

\textbf{MesoGraph}: In \cite{eastwood2023mesograph}, TIAViz is used to overlay subtype predictions onto mesothelioma TMA cores as can be seen at \url{https://tiademos.dcs.warwick.ac.uk/Bokeh_app?demo=mesograph}. The model illustrated here is a GNN on the cell graphs of the TMAs, using both morphological and local deep features to describe each cell.

\section{Discussion and Conclusions}

TIAViz is a powerful and flexible tool for visualization and exploration of digital pathology models. It is most useful in a research context where its browser-based nature allows use on remote machines and to provide online demos for collaborative purposes. As it is built in Python it also lends itself to quick integration, iteration and adaptation within the Python-based machine learning pipelines commonly used in digital pathology research.

There are some areas for improvement. One of these is performance optimization when handling very large datasets, a common challenge in digital pathology. The tool can be slow on initial loading in the case of large numbers of annotations, though it is generally quite responsive after this initial delay. We aim to improve this in the future. We also plan to widen the variety of tools in TIAToolbox that can be run directly within the interface. Currently, HoVer-Net can be used within the interface, but there is scope to widen the options there.

The possibility also exists to provide a visualization server allowing users to upload a slide and overlay to visualize. We currently have a prototype server for this which may be developed into a visualization service for the digital pathology community in the future.

\section{Acknowledgements}
\noindent This project was partially funded by CRUK-STFC Early Detection Innovation Award.
\noindent FM and ME acknowledge funding support from EPSRC Grant EP/W02909X/1 and Pathlake.

\small{
\bibliographystyle{abbrv}
\bibliography{refs}
}

\end{document}


\section*{Supplementary Material}

\begin{figure}[h]

 \centerline{\includegraphics[width=14cm]{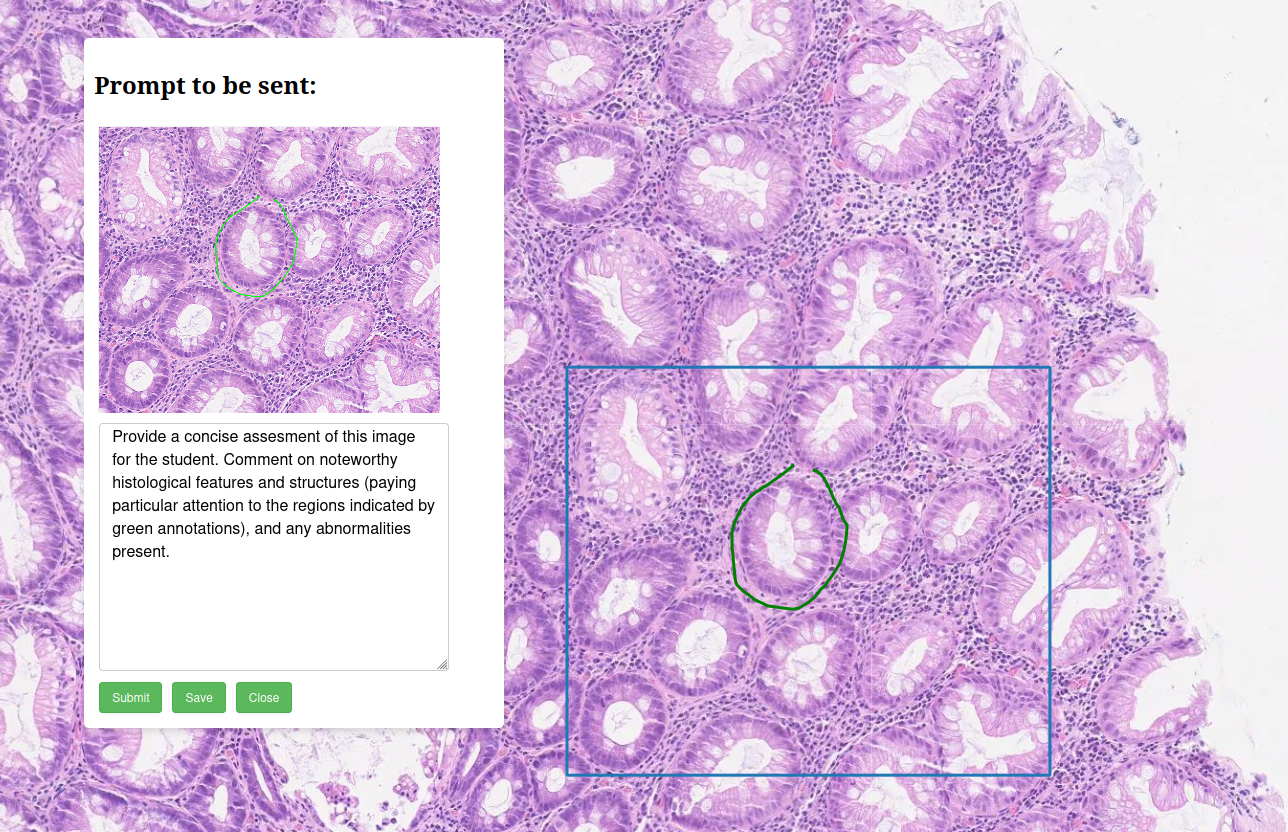}}

\caption{Example of a GPT-vision prompt sent from within the UI with added hand-drawn annotation. Response can be seen below.}
\label{gpt-vis}
\end{figure}

\textbf{GPT-vision response: }

This is an H\&E (hematoxylin and eosin) stained histological image that appears to show a section of the gastrointestinal tract, most likely the colon or small intestine due to the presence of goblet cells, which are mucus-secreting cells. The epithelial layer is columnar with regular nuclei, which is typical for this part of the digestive tract. The green annotation circles a gland in the epithelium.

What's noteworthy in this region:
\begin{enumerate}
\item 
In the annotated area, there is an intestinal gland (crypt of Lieberkühn), which looks to be lined with regularly spaced columnar epithelial cells with basal, round to oval nuclei. The presence of goblet cells interspersed between the absorptive cells is typical for normal intestinal mucosa.
\item 
Within the crypt, there is no evidence of dysplasia, which is characterized by variations in cell size (anisocytosis), shape (pleomorphism), nuclear enlargement, hyperchromasia, and increased mitotic activity. The cells retain uniformity and there's no apparent crowding or stratification of nuclei, which is reassuring.
\item 
The surrounding stroma contains lamina propria, which shows a moderate number of inflammatory cells which can be normal, but an increased number could indicate mild inflammation. However, there doesn't appear to be significant infiltrate to suggest active inflammation or an inflammatory bowel disease.
\item 
The crypt architecture is maintained and there doesn’t appear to be any evidence of cryptitis or crypt abscesses, which would suggest more specific inflammatory conditions such as ulcerative colitis.
\end{enumerate}

Abnormalities:
No overt abnormalities can be determined from this single region of the image, as the annotated crypt seems to have retained its normal histological architecture. Pathologists must, however, assess multiple fields and levels to establish a definite diagnosis and determine the absence of pathology accurately. 

For educational review, always correlate microscopic findings with macroscopic observations and clinical data for comprehensive pathology assessment.

\begin{figure*}

  \centering
 \centerline{\includegraphics[width=11cm]{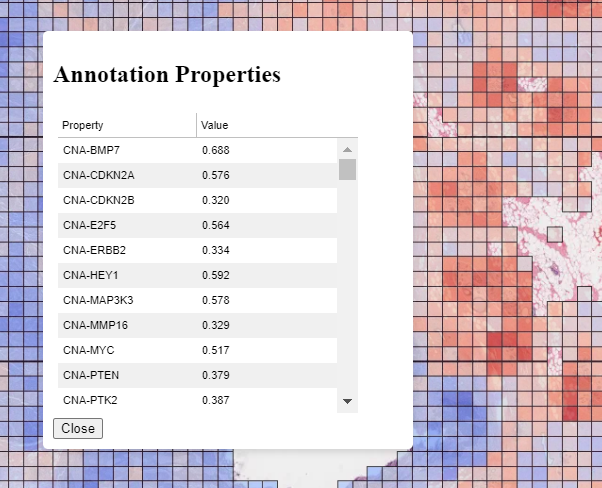}}

\caption{Double-clicking on an annotation will bring up a popup window giving a detailed table view of all properties of that annotation.}
\label{props_window}
\end{figure*}

\begin{figure*}

  \centering
 \centerline{\includegraphics[width=10cm]{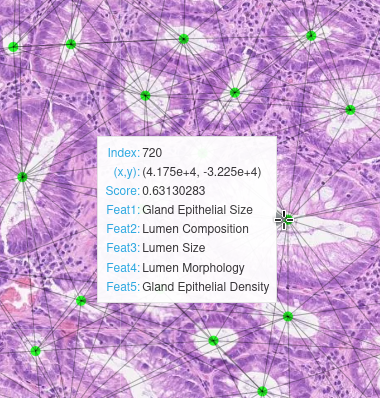}}

\caption{Hovering over a graph node will bring up further information about that node.}
\label{node_hover}
\end{figure*}